\documentclass[
 aps, pra,
 amsmath,amssymb,
 11pt,
 final,
tightenlines,
 twoside,
 twocolumn,
 nofloats,
nofootinbib,
 superscriptaddress,
showkeys,
showkeywords,
 ]
{revtex4-2}

\usepackage[T2A]{fontenc}
\usepackage[utf8x]{inputenc}
\usepackage[english]{babel}
\usepackage{graphicx}
\usepackage{dcolumn}
\usepackage{bm}
\usepackage{multirow} 
\usepackage{siunitx}

\def\squareforqed{\hbox{\rlap{$\sqcap$}$\sqcup$}}

\def\sq{\ifmmode\squareforqed\else{\unskip\nobreak\hfil
\penalty50\hskip1em\null\nobreak\hfil\squareforqed
\parfillskip=0pt\finalhyphendemerits=0\endgraf}\fi}

\def\arcsec{\hbox{$^{\prime\prime}$}}

\def\utw{\smash{\rlap{\lower5pt\hbox{$\sim$}}}}

\def\udtw{\smash{\rlap{\lower6pt\hbox{$\approx$}}}}

\def\fd{\hbox{$\,.\!\!^{\rm d}$}}

\def\fm{\hbox{$\,.\!\!^{\rm m}$}}

\def\diameter{{\ifmmode\mathchoice
{\ooalign{\hfil\hbox{$\displaystyle/$}\hfil\crcr
{\hbox{$\displaystyle\mathchar"20D$}}}}
{\ooalign{\hfil\hbox{$\textstyle/$}\hfil\crcr
{\hbox{$\textstyle\mathchar"20D$}}}}
{\ooalign{\hfil\hbox{$\scriptstyle/$}\hfil\crcr
{\hbox{$\scriptstyle\mathchar"20D$}}}}
{\ooalign{\hfil\hbox{$\scriptscriptstyle/$}\hfil\crcr
{\hbox{$\scriptscriptstyle\mathchar"20D$}}}}
\else{\ooalign{\hfil/\hfil\crcr\mathhexbox20D}}%
\fi}}

\setcitestyle{authoryear,aysep={,},yysep={;},semicolon,round}

\makeatletter
\renewcommand \thefigure{\@arabic\c@figure} 
\renewcommand \thetable{\@arabic\c@table} 

\usepackage{xurl} 
\usepackage[breaklinks]{hyperref}

\hypersetup{
    colorlinks=true,
    linkcolor=blue,
    filecolor=magenta,
    urlcolor=blue,
    citecolor=blue,
    breaklinks=true 
}

\begin{document}

\setlength{\tabcolsep}{0.5em} 

\selectlanguage{english}


\title{Orbital Parameters of the Unusual WR+O Binary System LS\,III\,+44\,21}

\author{\firstname{I.~I.}~\surname{Antokhin}}
\email{igor@sai.msu.ru}
\affiliation{Lomonosov Moscow State University, Sternberg State Anstonomical Institute, Moscow, Univeristitskij pr., 13, 119234, Russian Federation}

\author{\firstname{I.~A.}~\surname{Shaposhnikov}}
\affiliation{Lomonosov Moscow State University, Sternberg State Anstonomical Institute, Moscow, Univeristitskij pr., 13, 119234, Russian Federation}

\begin{abstract}

We present the results of a spectroscopic study of the recently discovered WR\,+\,O binary system LS\,III\,+44\,21. The system is unusual because, despite having characteristics similar to those of the classical WR\,+\,O system V444~Cyg, its X-ray emission is completely absent. We refined the spectral classification of the system components to \mbox{WN4\,+\,O7III?(f)} and obtained their radial velocity curves for the first time. The solution of these curves reveals that the system has a circular orbit. Using photometric observations from ASAS-SN and TESS, we significantly refined the values of the initial epoch $T_0$ and the orbital period $P$. The radial velocity curve solution with the updated $T_0$ and $P$ allowed us to determine the parameters of the orbit and the system components for the first time. A preliminary qualitative analysis of the light curves obtained by the TESS satellite and the ASAS-SN project, combined with the interesting variability of the N\,IV and N\,V ion line profiles with the orbital phase that we detected, suggests that the lack of X-ray emission from the system may be due to an unusually weak stellar wind from the WR star. The shape of the highly precise mean TESS light curve is extremely unusual, exhibiting non-monotonic behavior near the quadratures and an unusual morphology of the secondary minimum, which shows a nearly total eclipse yet with rounded edges. Such a shape cannot be modeled within the standard Roche geometry, which may provide a direct indication of the influence of the component stellar winds.

\end{abstract}

\maketitle

\section{Introduction}

LS\,III\,+44\,21 is a binary system that, until recently, managed to escape the close attention of astronomers despite its relative brightness ($V\simeq 10\fm 88$). Currently, the SIMBAD database contains only 15 bibliographic entries mentioning this object. The first mention of LS\,III\,+44\,21 is found in the ``Luminous Stars in the Northern Milky Way'' survey, which cataloged more than 5000 objects \citep{hardorp64}. The object --- not yet recognized as a binary system --- was subsequently included in several catalogs and databases \citep{koh97, reed98, koh99, reed03, reed05, clarke05, bourg14, cruzal19}, which either provided new measurements (e.g., {\em UBV} photometry) or cross-correlated existing catalogs. For a long time, the object was classified simply as an ``$\rm H_\alpha$ emission star.'' It was also observed by the Gaia mission, and its parallax, along with those of other stars, was utilized in studies mapping the spiral structure of the Milky Way in the solar neighborhood \citep{pant21, xu21}.

It was only in 2023, based on Gaia DR3 data, that the object was included in a list of eclipsing variable candidates \citep{movlavi23}. Using Gaia spectra, \cite{marin24} first identified the object as a binary system and determined the spectral classifications of its components to be WN6\,+\,O6.5\,V. The authors also incorporated photometric observations of the system obtained by the All-Sky Automated Survey for Supernovae \citep[ASAS-SN,][]{shappee14, hart23}. In the ASAS-SN database, the system is designated as \mbox{ASASSN-V~J204336.44+445505.8}. The $V$-band light curve exhibits a characteristic profile of an eclipsing binary system, with primary and secondary minimum depths of approximately $0\fm18$ and $0\fm13$, respectively, and an orbital period of $P\sim 4\fd43$. The secondary eclipse occurs exactly at orbital phase $0.5$, indicating a circular orbit. Recently, \cite{mulato25} confirmed the spectral classification proposed by \cite{marin24}.

Until now, a radial velocity curve and, consequently, the orbital parameters of the system have been lacking. Nevertheless, even the aforementioned information is sufficient to note that LS\,III\,+44\,21 appears remarkably similar to V444~Cyg, a system widely considered a benchmark for studying and understanding the physics of WR\,+\,O binaries. Indeed, V444~Cyg has an orbital period of $P\simeq4\fd2$ and a circular orbit. The spectral types of its components are WN5\,+\,O6\,V. The depths of the primary and secondary minima in the optical light curve of V444~Cyg are $\sim0\fm3$ and $0\fm14$, respectively. Comparing these depths with those of LS\,III\,+44\,21 suggests that the WR/O luminosity ratio in the latter is somewhat closer to unity than in V444~Cyg. In all other respects, however, the two systems appear to be close analogues.

\begin{figure*}
\includegraphics[width=0.45\textwidth]{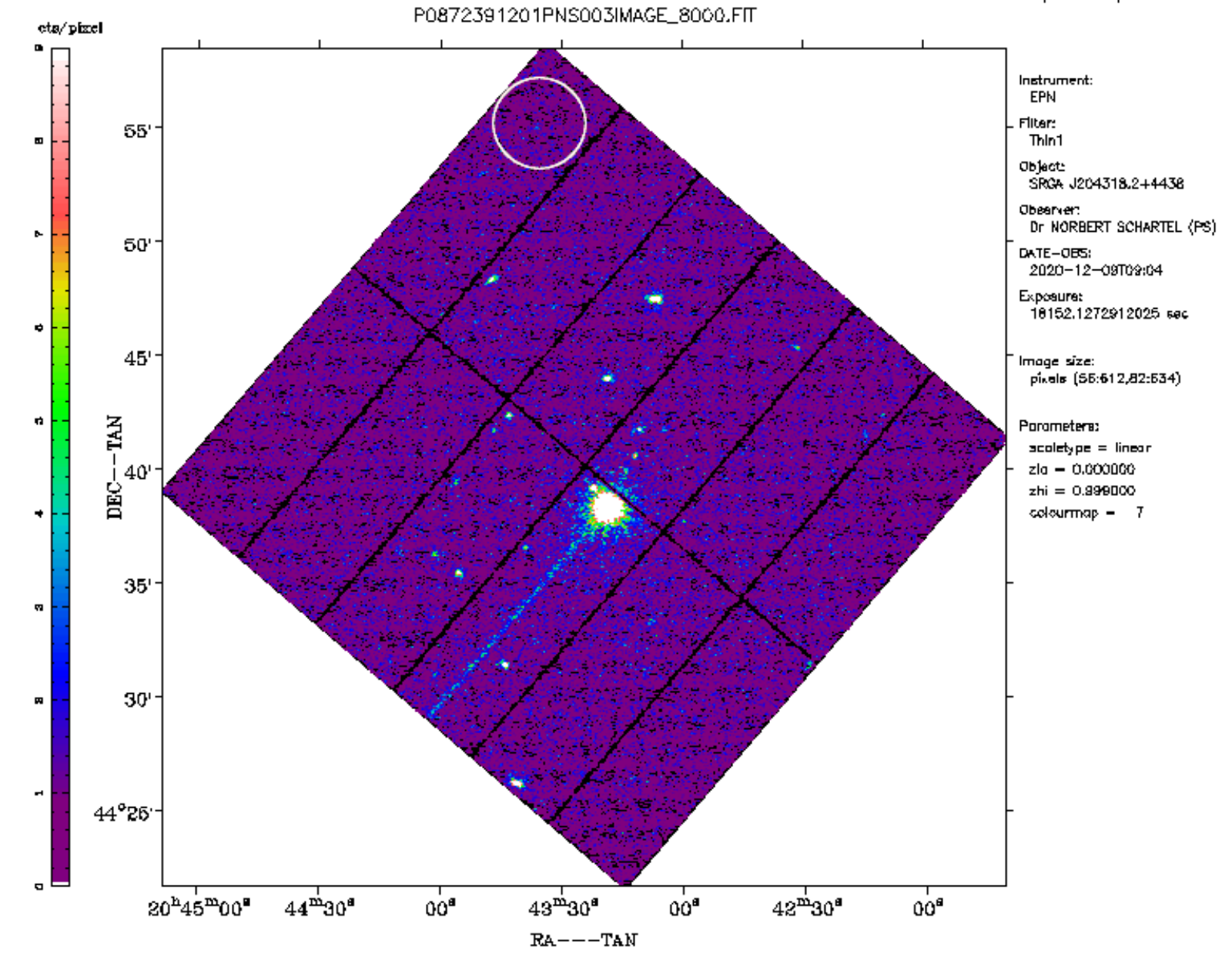}
\includegraphics[width=0.45\textwidth]{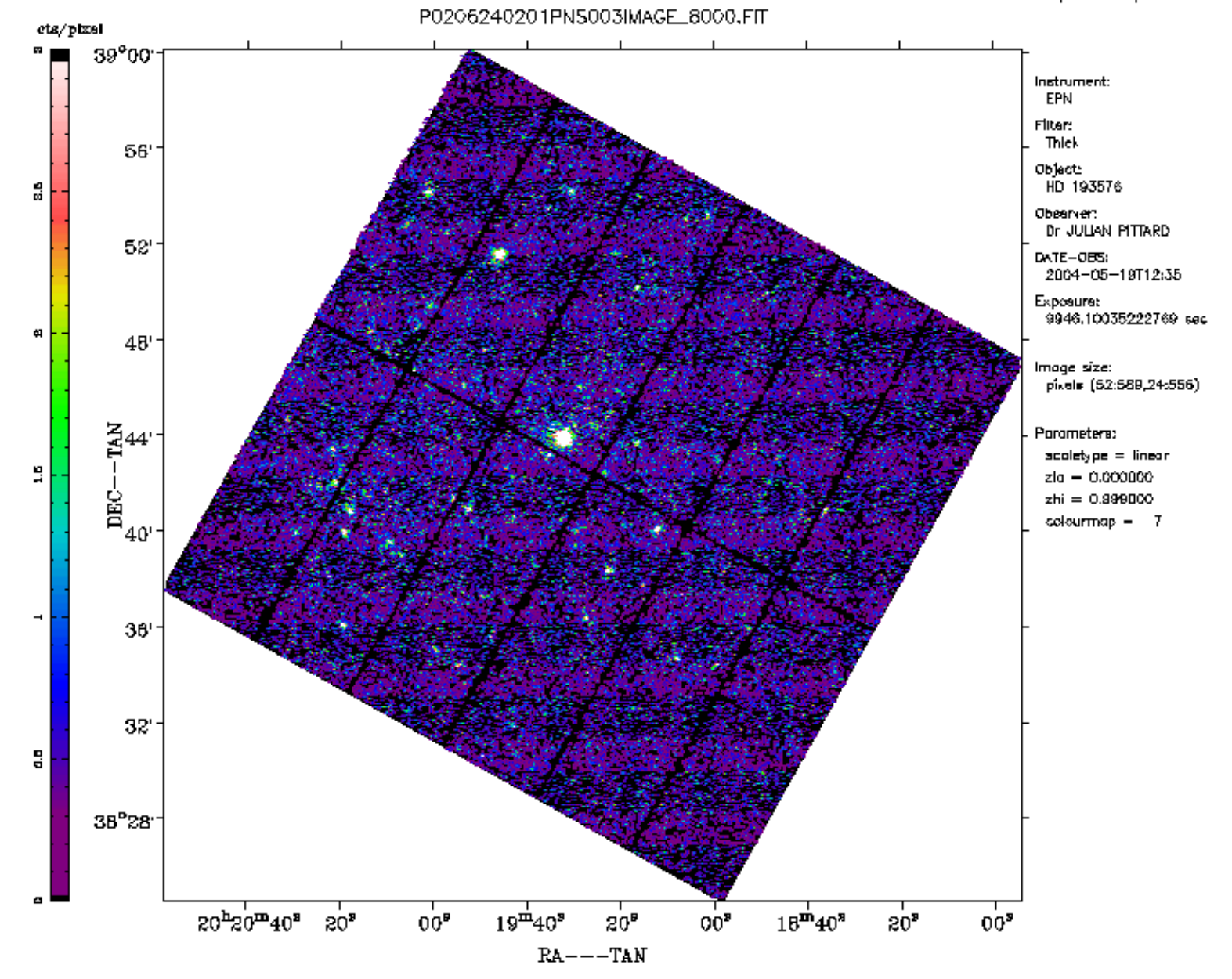}
\caption{Left: An image of the field around LS\,III\,+44\,21 obtained with the EPIC PN detector onboard the {\em XMM-Newton} telescope, with the position of the system indicated by a white circle. Right: A similar image for V444~Cyg, where the system is the brightest object in the center of the frame.}
\label{fig:xmm_image}
\end{figure*}

A distinctive feature of WR\,+\,O systems is the collision of the supersonic stellar winds from their components. The wind material within the shock region is heated to extremely high temperatures, resulting in strong and hard X-ray emission. This effect was predicted in the pioneering theoretical works by \cite{prilusov76} and \cite{cher76}, where it was proposed as a key criterion for searching for binary systems among Wolf-Rayet stars. As has become clearer in recent years, however, the mechanism behind X-ray generation is more complex than previously assumed. In close binary systems, the hardness and luminosity of the X-ray emission can be significantly reduced by radiative inhibition and radiative braking effects, which occur when the radiation field of one component decelerates the stellar wind of the other \citep{owocki95, owocki07}. Nonetheless, V444~Cyg remains a well-known X-ray source whose X-ray flux modulates with the orbital phase.

In stark contrast, there are no traces of X-ray emission from LS\,III\,+44\,21. As an illustration, Figure~\ref{fig:xmm_image} shows an image of the field around the system obtained with the EPIC PN detector onboard the {\em XMM-Newton} X-ray space observatory. No signs of an X-ray source are detected at the coordinates of LS\,III\,+44\,21. Evidently, the lack of X-ray emission is the primary reason the system remained unidentified as a WR\,+\,O binary for so long. Using the High Energy Astrophysics Science Archive Research Center (HEASARC) archive, we cross-checked data from other X-ray missions. According to {\em ROSAT} and {\em RXTE} observations, there are no indications of X-ray emission at the position of the system. The field containing the system was not observed by the {\em Chandra} or {\em eROSITA} satellites.

The total absence of X-ray emission from this system is highly unusual. Why does the very similar system V444~Cyg act as an X-ray source, while LS\,III\,+44\,21 does not? To clarify the origin of this discrepancy, it is essential to gather as much observational constraints on the system as possible. The fundamental information required is the set of orbital parameters, which can be derived from the radial velocity curves of the components. Therefore, the primary objective of this work is to present and analyze spectroscopic observations of the system.

The description of the observations and data reduction is provided in Section~\ref{sec:obs_red}. The obtained spectra and the results of the radial velocity curve analysis are presented in Section~\ref{sec:spec_rv}. In addition to analyzing the radial velocity curves, we utilized the light curves of LS\,III\,+44\,21 obtained by the ASAS-SN project and the TESS (Transiting Exoplanet Survey Satellite) mission to refine the initial epoch and the orbital period. Section~\ref{sec:phys_pars} provides a discussion on the physical parameters of the system and its individual components.
Finally, the main conclusions of this study are summarized in Section~\ref{sec:concl}.

\section{Observations and data reduction}\label{sec:obs_red}

Spectroscopic observations of the LS\,III\,+44\,21 system were carried out from December 2024 to January 2026 using the 2.5-meter telescope at the Caucasus Mountain Observatory of Lomonosov Moscow State University equipped with the TDS low-resolution double-beam spectrograph \citep{tds}. Measurements were performed simultaneously in two channels (``blue'' and ``red'') covering an effective wavelength range of $3600-7400$\,\AA\ with an overlap near $5700$\,\AA. The spectral resolution of the instrument in the long-slit mode with a $1\arcsec$ slit width is $R=1300$ in the ``blue'' channel and $R=2500$ in the ``red'' channel. The data reduction pipeline includes bias correction, dark current subtraction, and flat-fielding. Flat-field calibration frames were obtained for each telescope pointing using a continuum lamp. The dispersion curve was constructed for each row of the frame based on the emission-line spectrum of a Ne-Pb gas-discharge lamp.

Observations of LS\,III\,+44\,21 were conducted using a $1\arcsec$ long slit. For each telescope pointing, a series of three individual frames with exposure times of 300--450\,s per frame was obtained and subsequently averaged. The long total exposure time allowed us to record the background sky emission spectrum, the lines of which were used for an additional correction of the dispersion curve. On several nights with high seeing quality ($\approx1\arcsec$), the object was slightly moved along the slit by a small amount ($1-2\arcsec$) during the exposure to avoid detector saturation in the regions of strong spectral lines. Corrections for the spectral sensitivity of the detector, atmospheric extinction, and instrumental throughput were performed using spectra of A0\,V standard stars observed at a similar airmass during each pointing. In total, 35 spectra of LS\,III\,+44\,21 were obtained, providing good phase coverage of the expected orbital period of the system.

\section{The Spectrum of LS\,III\,+44\,21 and the Radial Velocity Curve Solution}\label{sec:spec_rv}

\begin{figure*}
\includegraphics[width=\textwidth]{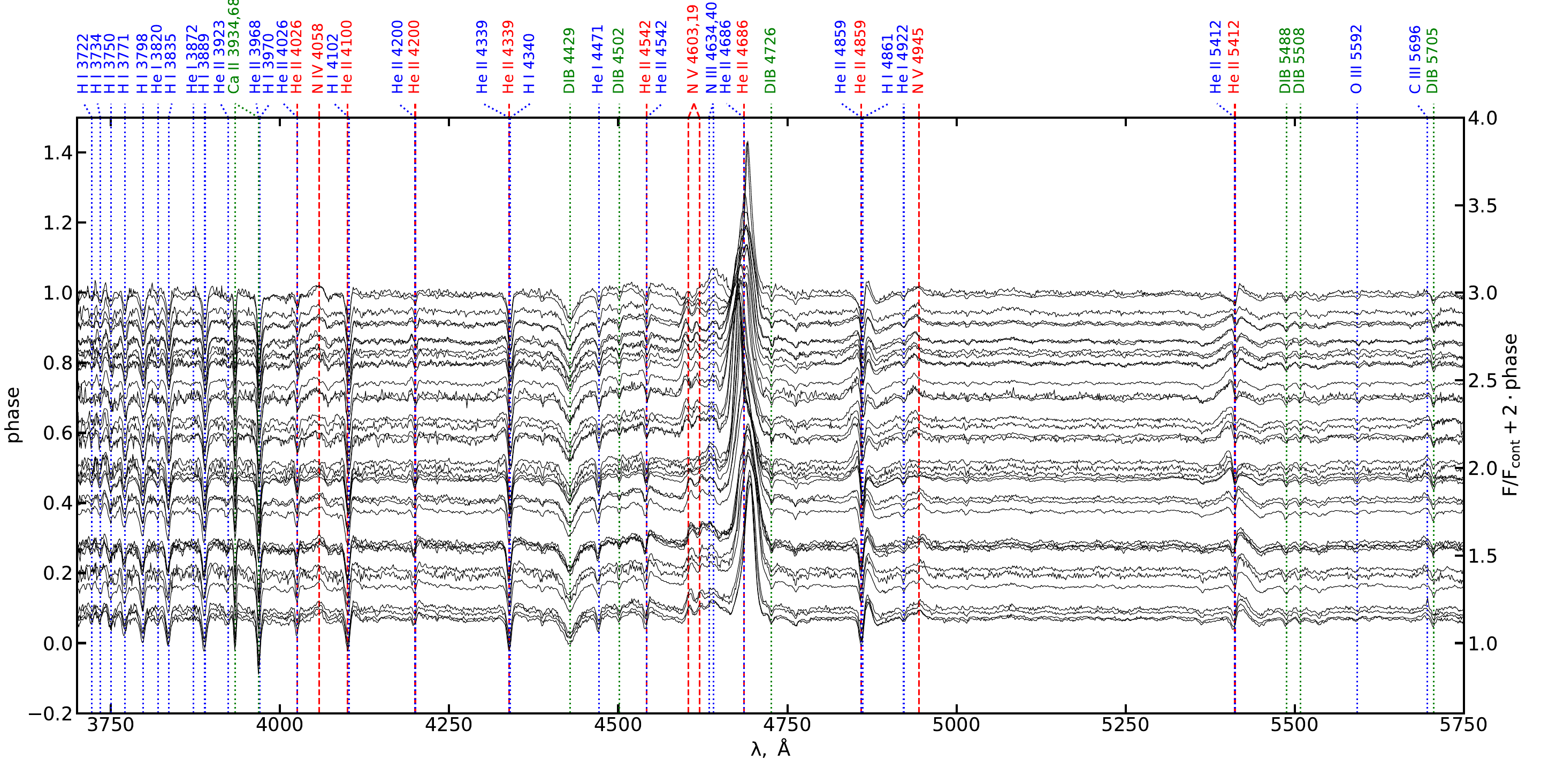}
\caption{Spectra of LS\,III\,+44\,21 obtained with the TDS spectrograph on the 2.5-meter telescope of the CMO SAI MSU, in the $3700-5750$\,\AA\ wavelength range (the ``blue'' channel of the instrument). The continuum level of each spectrum is scaled to the orbital phase, and the amplitude of the spectra is reduced by a factor of two. Blue color indicates features belonging to the O star, red indicates those of the WR star, and green represents interstellar features.}
\label{fig:sp-ident_B}
\end{figure*}

\begin{figure*}
\includegraphics[width=\textwidth]{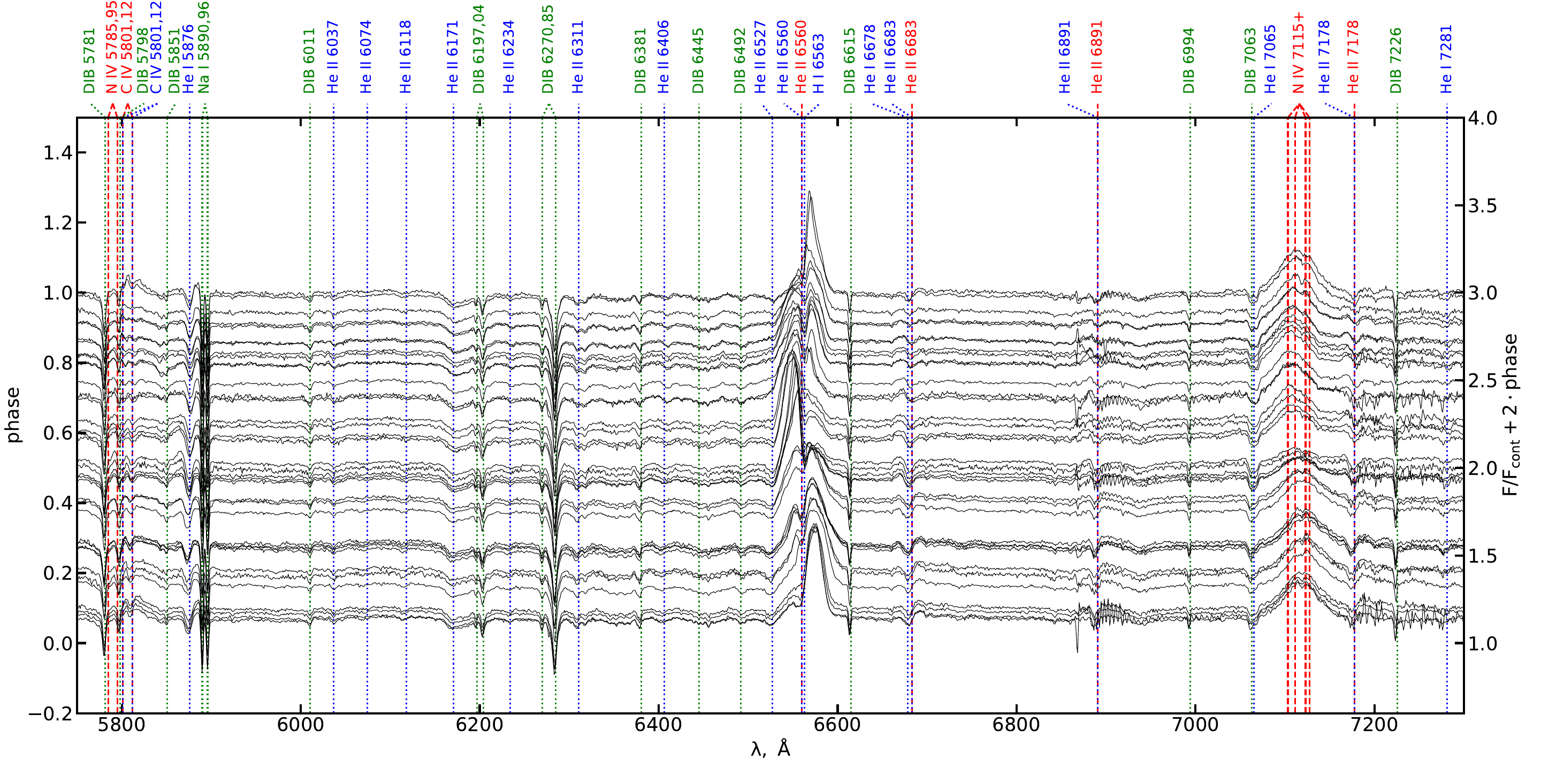}
\caption{Same as Figure ~\ref{fig:sp-ident_B}, but for the $5750-7300$\,\AA\ wavelength range (the ``red'' channel of the instrument).}
\label{fig:sp-ident_R}
\end{figure*}

\subsection{The Spectrum of LS\,III\,+44\,21 and Its Variability}\label{sec:spectrum}

The obtained spectra of the system are shown in Figures~\ref{fig:sp-ident_B} and \ref{fig:sp-ident_R}. The emission component of the spectra is represented by He\,II, N\,IV, and N\,V lines forming in the extended atmosphere of the WR star, as well as a blend of lines around $4640$\,\AA, which most likely corresponds to N\,III emissions. The latter emission group shifts in anti-phase with the others, allowing us to attribute it to the spectrum of the O star. The absorption features belong to the spectrum of the O star (H\,I, He\,I, He\,II, C\,IV, O\,III, and Si\,IV absorption lines) and to the interstellar medium (Ca\,II and Na\,I doublets, along with broad diffuse interstellar bands). The He\,II emissions of the WR star are strongly blended with the corresponding absorptions of the O star.

The accumulated spectroscopic material allows us to provide a more reliable spectral classification of the stars in the LS\,III\,+44\,21 system than that presented in \cite{marin24} and \cite{mulato25} based on a few spectra. Judging by its appearance in the region of the \mbox{N\,V~4604-4620\,\AA\ }lines (lower panel of Figure~3 in the cited paper), the spectrum presented by \cite{marin24} was obtained near phase $1.0$. At this phase, the intensity ratio of the N\,III, N\,IV, and N\,V emission lines differs from that observed at quadratures. Furthermore, as noted above, the N\,III emissions do not belong to the WR spectrum.

Therefore, the classification of the LS\,III\,+44\,21 system as WN6\,+\,O6.5\,V needs to be revised. \cite{mulato25} confirmed the classification of \cite{marin24} based on a single spectrum obtained with a \mbox{20-cm} telescope equipped with an extremely low-resolution spectrograph ($\sim 3$\,\AA/pixel), rendering their result equally unreliable. In our spectra obtained near orbital phases $0.25$ and $0.75$, when the features belonging to the O and WR spectra are not attenuated and are maximally separated, the WN stellar spectral classification criteria \cite{wrcat7} allow us to assign the WR star to the WN4 spectral subclass: the N\,III lines in its spectrum are weak or absent, and the intensities of the N\,IV $4058$\,\AA\ and \mbox{N\,V~$4604-4620$\,\AA\ }lines are comparable.

Based on the available spectral data, the O star can be classified as \mbox{O7\,III?(f)}. The spectral subclass $7$ appears preferable to $6.5$ because the intensities of the He\,I~4471\,\AA\ and He\,II~4541\,\AA\ lines look close, even taking into account the emission background near the latter, whereas for subclass $6.5$, the He\,II line should be slightly stronger. Unfortunately, the exact luminosity class of the O star is difficult to determine because the He\,II~4686\,\AA\ absorption used for classification is blended with the strong WR emission, and it is not possible to separate them correctly due to the low spectral resolution. The variability of the He\,II~4886\,\AA\ line profile width (wider at quadratures than at conjunctions) may also indicate the presence of an emission component of He\,II~4886\,\AA\ in the spectrum of the O star, which points toward a high luminosity (classes II--I). However, this criterion cannot be considered reliable, since the orbital variability of the He\,II~4886\,\AA\ line profile can be modulated by matter flows in the system. An indirect sign of the high luminosity of the O star in the LS\,III\,+44\,21 system is also the prominent nature of the absorption features compared to the WR emissions: most of the He\,II emissions (except for the strongest $4686$ and $6560$\,\AA\ lines) are almost completely suppressed by the absorption lines. \cite{marin24} assigned the O star to dwarfs (luminosity class V) based on a comparison of the intensity of the Si\,IV~4089 and 4116\,\AA\ lines with the H$_\delta$~4101\,\AA\ absorption. The indicated Si\,IV lines are barely discernible in our spectra. It should be noted that the criteria for assigning O stars to a particular spectral class based on the Si\,IV~4089 and 4116\,\AA\ lines were developed for late-subclass O stars (O9--O9.7, see, e.g., \citealt{martins18}), and \cite{marin24} do not specify a concrete criterion or provide any reference in their work. Thus, the luminosity class of the O star cannot yet be considered precisely determined, and the assignment of this star to giants (class III) represents a compromise based on a number of contradictory criteria.

\begin{table*}[t]
\centering
\scriptsize
\caption{Radial Velocities of LS\,III\,+44\,21 Derived from Different Lines (in $\rm km\,s^{-1}$)}

\label{tab:rv_obs}
\begin{tabular}{c S[separate-uncertainty=true] S[separate-uncertainty=true] S[separate-uncertainty=true] S[separate-uncertainty=true] S[separate-uncertainty=true] S[separate-uncertainty=true] S[separate-uncertainty=true]}
\noalign{\medskip}
\hline
\multicolumn{2}{c}{\vspace{-2.2ex}} \\
$BJD-2460000$ & $RV_{HI}^{3798}$ & $RV_{HI}^{3835}$ & $RV_{HI}^{3889}$ & $RV_{HeI}^{4471}$ & $RV_{HeI}^{5876}$ & $RV_{NV}^{4604}$ & $RV_{NIV}^{7117}$ \\
\multicolumn{2}{c}{\vspace{-2.4ex}} \\
\hline
651.293929  & -21 \pm 12 & 17 \pm 10 & 27 \pm 8 & 22 \pm 12 & 53 \pm 4 & -302 \pm 10 & -437 \pm 7 \\
652.156232  & -77 \pm 6 & -54 \pm 6 & -60 \pm 4 & -54 \pm 6 & -10 \pm 2 &  & -174 \pm 9 \\
654.258043  & -79 \pm 7 & -45 \pm 5 & -47 \pm 4 & -48 \pm 6 & -25 \pm 2 & 174 \pm 15 & -21 \pm 10 \\
655.286552  & 25 \pm 12 & 30 \pm 8 & 42 \pm 6 & 25 \pm 9 & 40 \pm 3 & -286 \pm 8 & -365 \pm 8 \\
784.539577  & 65 \pm 35 & 70 \pm 39 & 65 \pm 25 & 75 \pm 17 & 88 \pm 13 & -226 \pm 15 & -345 \pm 8 \\
785.478313  &  &  &  & -50 \pm 29 & -49 \pm 22 & 130 \pm 15 & -2 \pm 7 \\
887.531958  & -74 \pm 14 & -44 \pm 11 & -58 \pm 9 & -47 \pm 12 & -35 \pm 6 & 199 \pm 13 & 86 \pm 6 \\
888.313167  & -129 \pm 18 & -93 \pm 13 & -101 \pm 10 & -121 \pm 16 & -116 \pm 6 & 396 \pm 18 & 188 \pm 9 \\
891.275818  & -13 \pm 17 & 21 \pm 12 & -2 \pm 10 & -6 \pm 16 & 42 \pm 9 & -112 \pm 19 & -209 \pm 8 \\
898.535427  & 7 \pm 22 & -12 \pm 18 & 35 \pm 15 & 1 \pm 22 & 2 \pm 6 & -128 \pm 27 & -291 \pm 10 \\
899.492213  & 62 \pm 15 & 49 \pm 13 & 40 \pm 9 & -0 \pm 15 & 70 \pm 4 & -274 \pm 15 & -416 \pm 8 \\
902.521704  & -49 \pm 17 & -39 \pm 12 & -39 \pm 9 & -79 \pm 12 & -34 \pm 4 &  & -97 \pm 13 \\
903.509337  & 48 \pm 33 & 68 \pm 26 & 67 \pm 16 & 9 \pm 19 & 78 \pm 7 & -235 \pm 25 & -379 \pm 10 \\
904.438307  & -27 \pm 11 & 8 \pm 7 & 3 \pm 6 & 5 \pm 10 & 27 \pm 4 & -145 \pm 11 & -247 \pm 8 \\
906.482065  & -127 \pm 9 & -109 \pm 6 & -122 \pm 5 & -123 \pm 7 & -110 \pm 3 & 267 \pm 8 & 133 \pm 7 \\
907.559852  & -113 \pm 22 & 22 \pm 15 & -12 \pm 14 & 2 \pm 20 & -3 \pm 6 & -213 \pm 16 & -391 \pm 10 \\
911.539063  & -19 \pm 11 & 5 \pm 9 & 8 \pm 7 & -10 \pm 9 & -23 \pm 3 &  &  \\
915.468099  & -104 \pm 9 & -83 \pm 7 & -98 \pm 6 & -99 \pm 8 & -95 \pm 4 & 196 \pm 10 & 63 \pm 6 \\
917.500853  & 16 \pm 12 & 25 \pm 8 & 27 \pm 7 & -9 \pm 12 & 43 \pm 4 & -233 \pm 9 & -342 \pm 6 \\
918.501596  & -133 \pm 14 & -76 \pm 8 & -123 \pm 7 & -90 \pm 10 & -55 \pm 3 & 177 \pm 10 & 56 \pm 7 \\
925.370085  & 17 \pm 12 & 48 \pm 8 & 16 \pm 8 & 22 \pm 11 & 65 \pm 4 & -196 \pm 10 & -404 \pm 10 \\
979.418986  & -6 \pm 17 & 48 \pm 14 & 33 \pm 10 & 3 \pm 13 & 48 \pm 5 & -249 \pm 13 & -382 \pm 7 \\
981.391230  & -164 \pm 21 & -122 \pm 18 & -135 \pm 14 & -187 \pm 23 & -131 \pm 6 & 358 \pm 17 & 192 \pm 9 \\
982.251051  & -83 \pm 7 & -59 \pm 5 & -52 \pm 4 & -78 \pm 6 & -57 \pm 2 &  & -45 \pm 10 \\
984.189998  & 22 \pm 17 & 52 \pm 11 & 41 \pm 10 & 16 \pm 15 & 37 \pm 9 & -155 \pm 10 & -232 \pm 10 \\
985.305311  & -101 \pm 17 & -78 \pm 16 & -86 \pm 13 & -94 \pm 14 & -92 \pm 10 & 333 \pm 12 & 177 \pm 8 \\
987.214067  & 14 \pm 8 & 43 \pm 6 & 46 \pm 5 & 44 \pm 6 & 40 \pm 2 & -137 \pm 8 & -321 \pm 8 \\
991.232893  & -93 \pm 20 & -66 \pm 15 & -103 \pm 11 & -14 \pm 16 & -33 \pm 6 &  & -134 \pm 21 \\
994.312445  & -104 \pm 18 & -124 \pm 28 & -131 \pm 21 & -118 \pm 28 & -81 \pm 9 & 335 \pm 29 & 184 \pm 10 \\
998.189594  & -108 \pm 11 & -84 \pm 9 & -107 \pm 6 & -102 \pm 11 & -68 \pm 5 & 119 \pm 11 & 44 \pm 8 \\
999.161878  & -123 \pm 13 & -102 \pm 10 & -105 \pm 8 & -123 \pm 12 & -120 \pm 6 &  & 160 \pm 8 \\
1001.174857 & 10 \pm 8 & 45 \pm 6 & 43 \pm 5 & 30 \pm 8 & 68 \pm 3 & -303 \pm 8 & -428 \pm 7 \\
1002.318844 & -22 \pm 20 & -42 \pm 17 & -22 \pm 13 & -9 \pm 18 & 8 \pm 7 & -14 \pm 30 & -119 \pm 6 \\
1003.255194 & -136 \pm 14 & -116 \pm 12 & -129 \pm 10 & -120 \pm 14 & -129 \pm 11 & 334 \pm 15 & 198 \pm 10 \\
1004.152409 & -82 \pm 13 & -57 \pm 11 & -90 \pm 8 & -111 \pm 12 & -93 \pm 5 & 186 \pm 14 & 50 \pm 9 \\
1028.123294 & 58 \pm 14 & 62 \pm 11 & 65 \pm 9 & 23 \pm 15 & 47 \pm 5 & -260 \pm 13 & -418 \pm 7 \\
1061.124270 & -138 \pm 19 & -126 \pm 14 & -106 \pm 10 & -135 \pm 13 & -137 \pm 5 & 342 \pm 15 & 203 \pm 8 \\
\hline
\end{tabular}
\end{table*}

\subsection{Radial Velocity Measurements}\label{sec:rv_meas}

Radial velocity measurements for a number of spectral lines were performed by fitting a Gaussian profile along with a local linear continuum. The Gaussian function was used to approximate both the emission line profiles of the WR star and the absorption lines of the O star. Utilizing more complex model profiles, such as Lorentz or Voigt functions, does not yield higher precision compared to a Gaussian profile due to the low spectral resolution (\mbox{$R = 1300-2500$}).

To obtain information on the orbital motion of the O star, we selected the least blended and sufficiently strong absorption lines, which turned out to be the hydrogen Balmer series lines \mbox{H\,I~3798, 3835, and 3889\,\AA\ }and the neutral helium lines \mbox{He\,I~4471 and 5876\,\AA}. We consider the \mbox{N\,V~4604\,\AA\ }line as the primary source of information regarding the orbital motion of the WR star, since its formation region should lie closest to the core of the WR star. We do not use the \mbox{N\,V} 4620 and 4944\,\AA\ lines for radial velocity measurements due to their severe blending. Near phases $0.0\,(1.0)$ and $0.5$, when the N\,V ion lines in the spectrum of LS\,III\,+44\,21 are very weak, it is impossible to measure their positions. In addition to N\,V, we measured the position of the strong \mbox{N\,V~7115+\,\AA\ }blend (the laboratory central wavelength was taken to be that of one of the blend components, $\lambda = 7117.213$\,\AA). The measurement results are presented in Table~\ref{tab:rv_obs}.

It is necessary to note that there is a scatter in the systematic velocities ($\gamma$-velocities) derived from the radial velocity curves of individual lines from Table~\ref{tab:rv_obs}. This is particularly noticeable for the H\,I lines that are close in wavelength. In the case of measurements based on the emission lines of various ions in the WR spectrum, the differences in $\gamma$-velocities and even the appearance of a spurious orbital eccentricity can be related to the kinematics of the extended WR atmosphere. The photospheric lines of the O star should not display a significant scatter in $\gamma$-velocities, yet it is present in our data. In the blue region of the spectrum, this scatter may be caused by a less accurate wavelength calibration (with a systematic error of no more than 15--20\,km/s) due to the absence of strong telluric emission lines in this spectral range, which we use for additional correction of the dispersion curve (see Section~\ref{sec:obs_red}). As for the He\,I lines, the inaccuracy in their $\gamma$-velocities may stem from a weak underlying emission component, which leads to a slight distortion of the profile and a shift of its apparent center. Furthermore, systematic errors in radial velocity measurements can arise from line profile distortions caused by radiation scattering as it passes through moving gas streams in the system. These effects can introduce a certain asymmetry into the real line profiles, though numerically estimating the degree of their influence remains difficult due to strong instrumental broadening. However, as shown below, the shapes and semi-amplitudes of the radial velocity curves for different lines are in good agreement with each other and thus reliably reflect the orbital motion of the system components. Given these considerations, in subsequent calculations we treat the $\gamma$-velocity for each obtained radial velocity curve as an independent model parameter.

\subsection{Radial Velocity Curve Solution}\label{sec:rv_sol}

We solved the radial velocity curves of LS\,III\,+44\,21 in two steps. In the first step, we employed the JADE algorithm (Adaptive Differential Evolution with Optional External Archive, \citealp{zhang2009jade}) to search for the global minimum of the $\chi^2$ residual. In the second step, we used the Markov Chain Monte Carlo (MCMC) algorithm \citep{metropolis53, hastings70}. To accelerate the performance of the MCMC algorithm, we incorporated a mechanism of speculative moves \citep{byrd2008speculative}. Both algorithms are implemented in C++ with multi-threading support; their detailed description and a link to a publicly available repository containing the source code are provided in \cite{ant26}.

The population size in the JADE algorithm was set to 5000, with 50 generations. The best individual from the final JADE population was chosen as the starting point for the Markov chain. In addition, the final JADE population was used to compute the covariance matrix, which allows evaluating potential correlations between the model parameters. This matrix was subsequently utilized in the MCMC algorithm to define the optimal shape and scale of the random steps (the proposal distribution) based on the already identified data structure. This approach enables the algorithm to efficiently explore the target distribution, bypassing a lengthy burn-in period for parameter tuning and radically increasing the chain efficiency. For each modeling run, the number of MCMC steps was set to 1\,000\,000.

\begin{figure*}
\includegraphics[width=\textwidth]{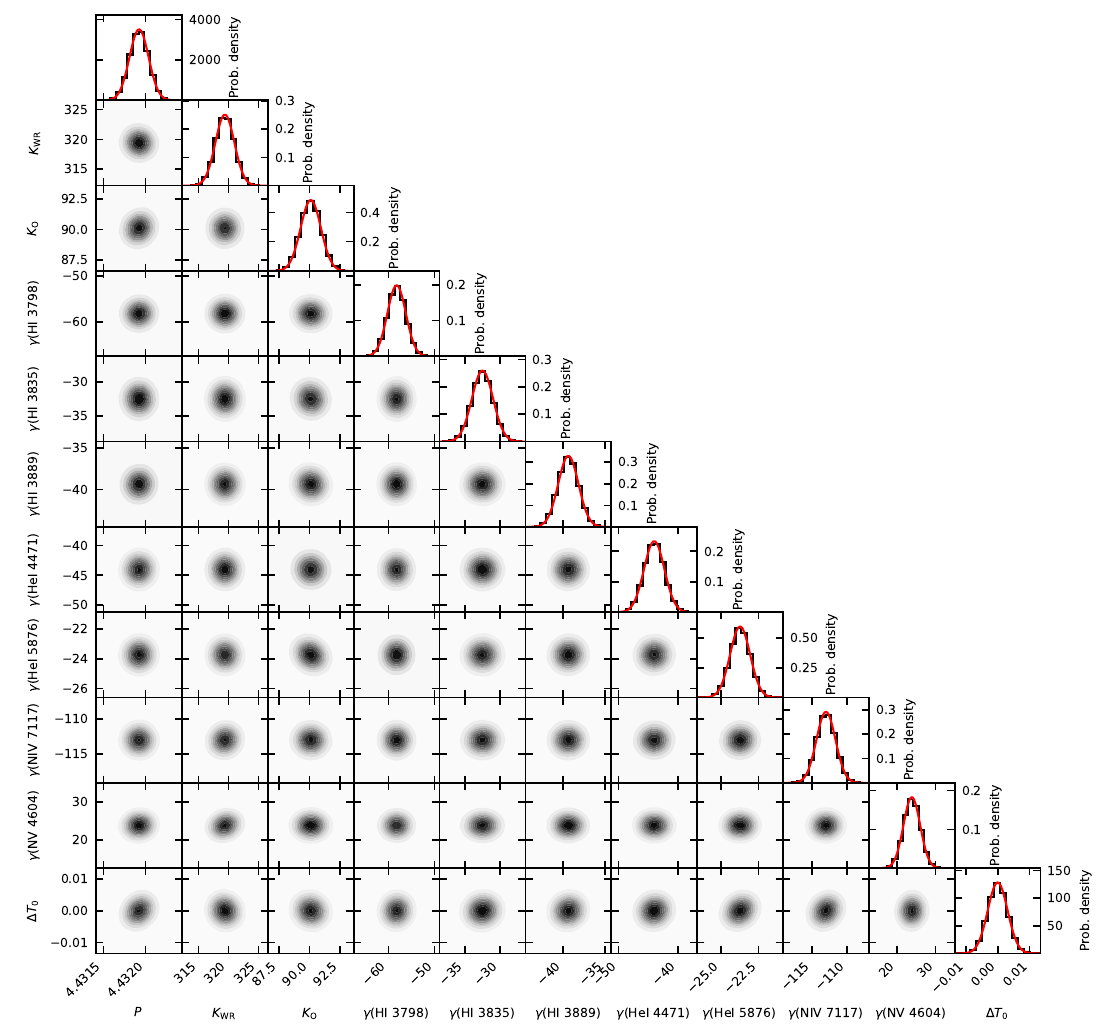}
\caption{Posterior empirical distributions for all possible pairs of model parameters. To save space, the differences between the initial epoch and its optimal value, $\Delta T_0$, are shown instead of the initial epoch $T_0$ itself. Actual values are displayed for all other parameters. On the right of each row (and on the top panel of the first column for the orbital period), the empirical probability density histograms for the corresponding parameter are shown along with their fitting Gaussian functions (indicated by solid red curves).}
\label{fig:corner_plot}
\end{figure*}

We solved the radial velocity curves for both circular and eccentric orbit configurations. In the latter case, the resulting orbital eccentricity was indistinguishable from zero within the errors. Therefore, the solution for a circular orbit is presented below. The optimal values of the model parameters are listed in Table~\ref{tab:rv_sol1}. Figure~\ref{fig:corner_plot} displays the posterior empirical distributions for all possible pairs of parameters. These distributions were used to compute the parameter uncertainties provided in Table~\ref{tab:rv_sol1}.

\begin{table}[t]
\centering
\footnotesize
\caption{Optimal Parameters of the Radial Velocity Curve Model Derived from the Spectral Data in This Work}
\label{tab:rv_sol1}
\begin{tabular}{l r @{\hskip 0.1em $\pm$ \hskip -0.02em} l}
\hline
& \multicolumn{2}{c}{} \\[-2.2ex]
Parameter & \multicolumn{2}{c}{Value} \\[0.8ex]
\hline
& \multicolumn{2}{c}{} \\[-2.2ex]
$T_0$ (BJD${}-2\,460\,000$)           & 913.682  & 0.003 \\
$P$ (days)                            & 4.43192  & 0.00011 \\
$K_{\rm WR}$ ($\rm km\,s^{-1}$)       & 319.4    & 1.6 \\
$K_{\rm O}$ ($\rm km\,s^{-1}$)        & 90.1     & 0.8 \\
$\gamma_{\rm O}$ (H\,I 3798, km\,s$^{-1}$)   & -58.2    & 2.0 \\
$\gamma_{\rm O}$ (H\,I 3835, km\,s$^{-1}$)   & -32.5    & 1.5 \\
$\gamma_{\rm O}$ (H\,I 3889, km\,s$^{-1}$)   & -39.4    & 1.2 \\
$\gamma_{\rm O}$ (He\,I 4471, km\,s$^{-1}$)  & -44.0    & 1.7 \\
$\gamma_{\rm O}$ (He\,I 5876, km\,s$^{-1}$)  & -23.7    & 0.7 \\
$\gamma_{\rm WR}$ (N\,IV 7117, km\,s$^{-1}$) & -113.0   & 1.4 \\
$\gamma_{\rm WR}$ (N\,V 4604, km\,s$^{-1}$)  & 23.8     & 2.2 \\[0.5ex]
\hline
\end{tabular}
\end{table}

\subsection{Refinement of the Initial Epoch and Orbital Period Using {\rm ASAS-SN} and {\rm TESS} Light Curves}\label{sec:asas_tess}

\begin{table*}[!t]
\centering
\scriptsize
\caption{TESS light-curve files of LS\,III\,+44\,21 used in this work. Listed are the sector numbers, the start and end dates of observations for each sector, the number of brightness measurements, and the data source. The varying number of measurements across different sectors reflects different binning intervals of individual exposures.}
\label{tab:tess_data}
\begin{tabular}{lcccc}
\noalign{\medskip}
\hline
\multicolumn{5}{c}{\vspace{-2.0ex}} \\
\multicolumn{1}{c}{File name} & \multicolumn{1}{c}{Sector} & Interval   & \multicolumn{1}{c}{N} & \multicolumn{1}{c}{Source} \\
\multicolumn{5}{c}{\vspace{-2.0ex}} \\
\hline
\verb|hlsp_qlp_tess_ffi_s0015-0000000196054579_tess_v01_llc.fits|      & 15 & 15.08--10.09.2019 &  1197 & QLP  \\
\verb|hlsp_qlp_tess_ffi_s0041-0000000196054579_tess_v01_llc.fits|      & 41 & 23.07--20.08.2021 &  3668 & QLP  \\
\verb|hlsp_qlp_tess_ffi_s0055-0000000196054579_tess_v01_llc.fits|      & 55 & 05.08--01.09.2022 &  3781 & QLP  \\
\verb|hlsp_tess-spoc_tess_phot_0000000196054579-s0056_tess_v1_lc.fits| & 56 & 01.09--30.09.2022 & 11764 & SPOC \\
\verb|hlsp_tess-spoc_tess_phot_0000000196054579-s0075_tess_v1_lc.fits| & 75 & 30.01--26.02.2024 & 11768 & SPOC \\
\verb|hlsp_qlp_tess_ffi_s0075-0000000196054579_tess_v02_llc.fits|      & 75 & 30.01--26.02.2024 & 11634 & QLP  \\
\verb|hlsp_qlp_tess_ffi_s0082-0000000196054579_tess_v01_llc.fits|      & 82 & 10.08--05.09.2024 & 10778 & QLP  \\
\verb|hlsp_qlp_tess_ffi_s0083-0000000196054579_tess_v01_llc.fits|      & 83 & 05.09--01.10.2024 & 10505 & QLP  \\
\hline
\end{tabular}
\end{table*}

LS\,III\,+44\,21 was observed in the $V$ band within the ASAS-SN project from March 2015 to July 2018, and by the TESS space telescope from 2019 August 15 to 2024 October 1. Our spectroscopic observations were carried out from December 2024 to January 2026. This extensive baseline provides an opportunity to refine the values of $T_0$ and $P$ found from our spectral data.

The ASAS-SN light curve consists of 129 individual measurements. On the object's page within the project database, there is a warning indicating that some measurements might be affected by saturation. However, the overall light curve closely resembles the TESS light curve (see below). Furthermore, since this work is exclusively concerned with the timings of the minima, we deemed it reasonable to include these data in our analysis.

The MAST (Barbara A. Mikulski Archive for Space Telescopes) database\footnote{\tt https://stsci.edu\\/Mast/Portal.html} contains observations of the system spanning seven TESS sectors. The list of the light-curve files used in this study is provided in Table~\ref{tab:tess_data}. The majority of these light curves originate from the QLP (Quick Look Pipeline, \citealp{kunimoto22}), and only two sectors were processed by the SPOC (Science Processing Operations Center). These two pipelines differ in their algorithms for photometric measurements, barycentric correction calculations, etc., which could theoretically introduce minor timing offsets (up to 3--6 minutes) between them. To test for this effect, we utilized data from both sources for Sector 75 (the only sector for which both pipelines are available) and verified that the timings of the minima derived from both sources are identical (see below). For Sector 75, we adopted the second version of the QLP light-curve file because, according to the MAST website, the first version contained incorrectly calculated barycentric corrections. Occasional bad data points that deviate sharply from the regular light curve are present in the TESS data. Such outliers (496 points in total across Sectors 15, 41, 55, and 83) were removed prior to using the data.

To determine the timings of the light-curve minima, we employed the Hertzsprung method in the modification proposed by \cite{kwee56}. The method was applied separately to the primary and secondary minima of a given light curve. Unfortunately, the total number of the ASAS-SN brightness measurements is too small to determine the timings of the minima for shorter time sub-intervals within the entire observational baseline. Therefore, we determined only a single time for the primary minimum and a single time for the secondary minimum, referencing them to the mean orbital cycle over the 2015--2018 interval, which automatically implies relatively large uncertainties. Using the TESS data, we determined one time for the primary minimum and one for the secondary minimum of the light curve for each individual sector (the average duration of a sector's observations is 27.4 days). These time are referenced to the mean orbital cycle within the observational interval of the respective sector. The list of the measured times of the light-curve minima across all data available to us is provided in Table~\ref{tab:jd_times}.

\begin{table}[!t]
\centering
\scriptsize
\caption{Times of the Light-Curve Minima of LS\,III\,+44\,21 Derived from ASAS-SN, TESS, and This Work. The first and second columns show the Barycentric Julian Date (BJD) and its uncertainty. The third column lists the number of the light-curve minimum (1 for primary, 2 for secondary) used to determine the given timing. For convenience, the times of the secondary minima are converted into the times of the primary ones. The last column indicates the data source and the sector number (for the TESS data).}
\label{tab:jd_times}
\begin{tabular}{cccc}
\noalign{\medskip}
\hline
\rule{0pt}{3ex} BJD      & $\sigma$(BJD) & $N_{min}$ & Source   \rule[-1.5ex]{0pt}{0pt}\\
\hline
 2457616.665 & 0.046 &  1   &  ASAS-SN \\
 2457616.677 & 0.029 &  2   &  ASAS-SN \\
 2458724.535 & 0.003 &  1   &   QLP 15 \\
 2458724.542 & 0.003 &  2   &   QLP 15 \\
 2459433.593 & 0.001 &  1   &   QLP 41 \\
 2459433.593 & 0.004 &  2   &   QLP 41 \\
 2459814.710 & 0.003 &  1   &   QLP 55 \\
 2459814.709 & 0.003 &  2   &   QLP 55 \\
 2459841.294 & 0.002 &  1   &  SPOC 56 \\
 2459841.297 & 0.003 &  2   &  SPOC 56 \\
 2460355.352 & 0.002 &  1   &   QLP 75 \\
 2460355.354 & 0.003 &  2   &   QLP 75 \\
 2460355.352 & 0.002 &  1   &  SPOC 75 \\
 2460355.354 & 0.003 &  2   &  SPOC 75 \\
 2460550.342 & 0.001 &  1   &   QLP 82 \\
 2460550.340 & 0.005 &  2   &   QLP 82 \\
 2460572.500 & 0.001 &  1   &   QLP 83 \\
 2460572.501 & 0.003 &  2   &   QLP 83 \\
 2460913.682 & 0.003 &  --  & Эта раб. \\
\hline
\end{tabular}
\end{table}

\begin{figure}
\includegraphics[width=\columnwidth]{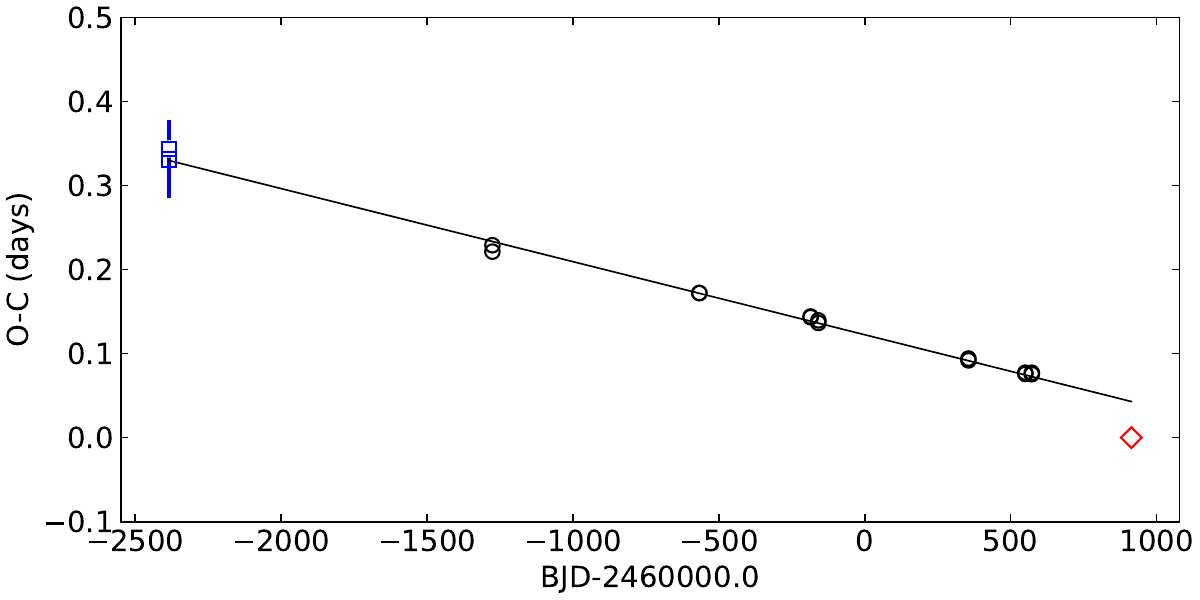}
\caption{Linear least-squares fit to the times of the light-curve minima. Shown are the residuals ($O-C$) of the observed times of minima relative to those calculated from the linear ephemeris using $T_0$ and $P$ from Table~\ref{tab:rv_sol1}. The two leftmost blue square points are derived from the ASAS-SN data. The rightmost red diamond represents the data from this work. The black open circles indicate the TESS data.}
\label{fig:oc_fit}
\end{figure}

Figure~\ref{fig:oc_fit} shows the results of a linear least-squares fit to the times of the light-curve minima,

$$
 T({\rm Min~I}) = T_0 + P\cdot n\,
$$

where $n$ is the number of orbital cycles elapsed since the initial epoch. During the fitting process for Sector 75, only the SPOC data from Table~\ref{tab:jd_times} were utilized.

The refined values of $T_0$ and $P$ along with their uncertainties, derived from the analysis of the times of minima, are listed in Table~\ref{tab:rv_sol2}. A comparison with the corresponding quantities in Table~\ref{tab:rv_sol1} demonstrates that incorporating the ASAS-SN and TESS light curves has drastically improved the precision of these parameters. For this reason, we repeated the radial velocity model optimization using the MCMC algorithm once more, fixing the values of $T_0$ and $P$ and leaving only $K_1$, $K_2$, and the $\gamma$-velocities of the measured lines as free parameters. The optimization results are summarized in Table~\ref{tab:rv_sol2}.


\begin{table}[t]
\centering
\caption{Final Parameters of the Optimal Model Using the Refined Values of $T_0$ and $P$.}
\label{tab:rv_sol2}
\begin{tabular}{l r @{\hskip 0.1em $\pm$ \hskip 0.02em} l}
\noalign{\medskip}
\hline
\rule{0pt}{3ex} Parameter & \multicolumn{2}{c}{Value} \rule[-1.5ex]{0pt}{0pt}\\
\hline
\multicolumn{3}{c}{Fixed Parameters} \\
\hline
$T_0$ (BJD-2\,460\,000)  & 913.72491  & 0.00097 \\
$P$ (days)               & 4.431535   & 0.000004 \\
\hline
\multicolumn{3}{c}{Free Parameters} \\
\hline
$K_{\rm WR}$ (km\,s$^{-1}$)                 & 317.7   & 1.6  \\
$K_{\rm O}$ (km\,s$^{-1}$)                  & 88.4    & 0.8   \\
$\gamma_{\rm O}$(H\,I~3798, km\,s$^{-1}$)   & -57.5   & 0.8 \\
$\gamma_{\rm O}$(H\,I~3835, km\,s$^{-1}$)   & -31.4   & 1.5 \\
$\gamma_{\rm O}$(H\,I~3889, km\,s$^{-1}$)   & -38.5   & 1.2 \\
$\gamma_{\rm O}$(He\,I~4471, km\,s$^{-1}$)  & -42.7   & 1.7 \\
$\gamma_{\rm O}$(He\,I~5876, km\,s$^{-1}$)  & -22.2   & 0.7 \\
$\gamma_{\rm WR}$(N\,IV~7117, km\,s$^{-1}$) & -110.0  & 1.4 \\
$\gamma_{\rm WR}$(N\,V~4604, km\,s$^{-1}$)  & 24.0    & 2.2 \\
\hline
\multicolumn{3}{c}{Derived Parameters} \\
\hline
$a_{\rm WR}\sin i$ ($R_\odot$)             & 27.8    & 0.14  \\
$a_{\rm O}\sin i$ ($R_\odot$)              & 7.74    & 0.07  \\
$(a_{\rm WR}+a_{\rm O})\sin i$ ($R_\odot$) & 35.5    & 0.16  \\
$M_{\rm WR}\sin^3 i$ ($M_\odot$)           & 6.69    & 0.19  \\
$M_{\rm O}\sin^3 i$ ($M_\odot$)            & 24.06   & 0.38 \\
$q=M_{\rm O}/M_{\rm WR}$                   & 3.59    & 0.12  \\
\hline
\end{tabular}
\end{table}

\begin{figure}
\includegraphics[width=\columnwidth]{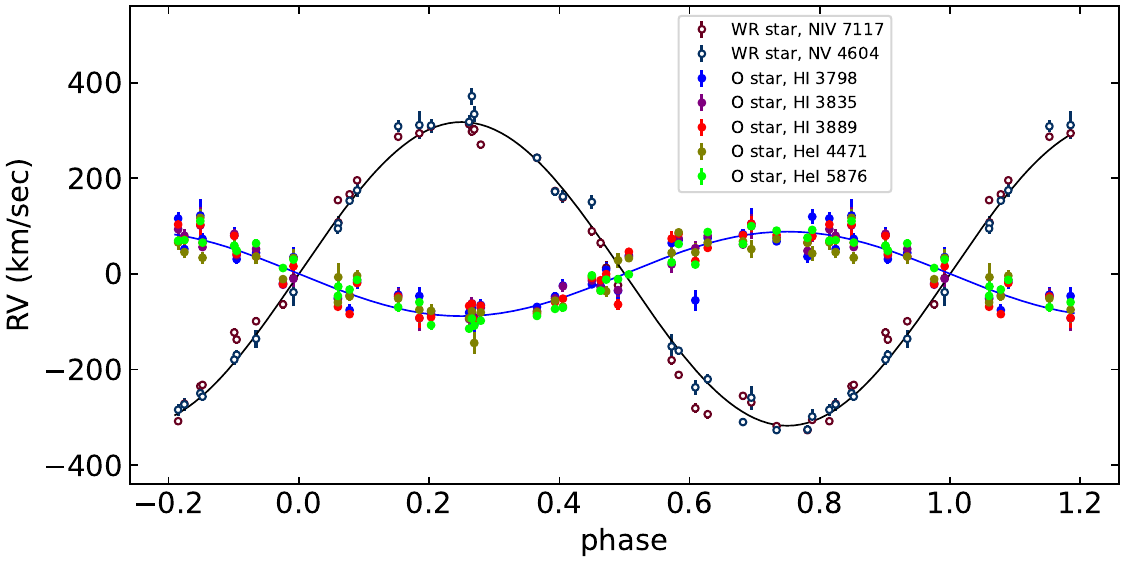}
\caption{Observational and theoretical radial velocity curves of the system components, shifted to the reference frame with a zero $\gamma$-velocity. The observed velocities of the WR star are shown as open circles, and those of the O star are displayed as filled circles.}
\label{fig:rv_fit}
\end{figure}

\begin{figure}
\includegraphics[width=\columnwidth]{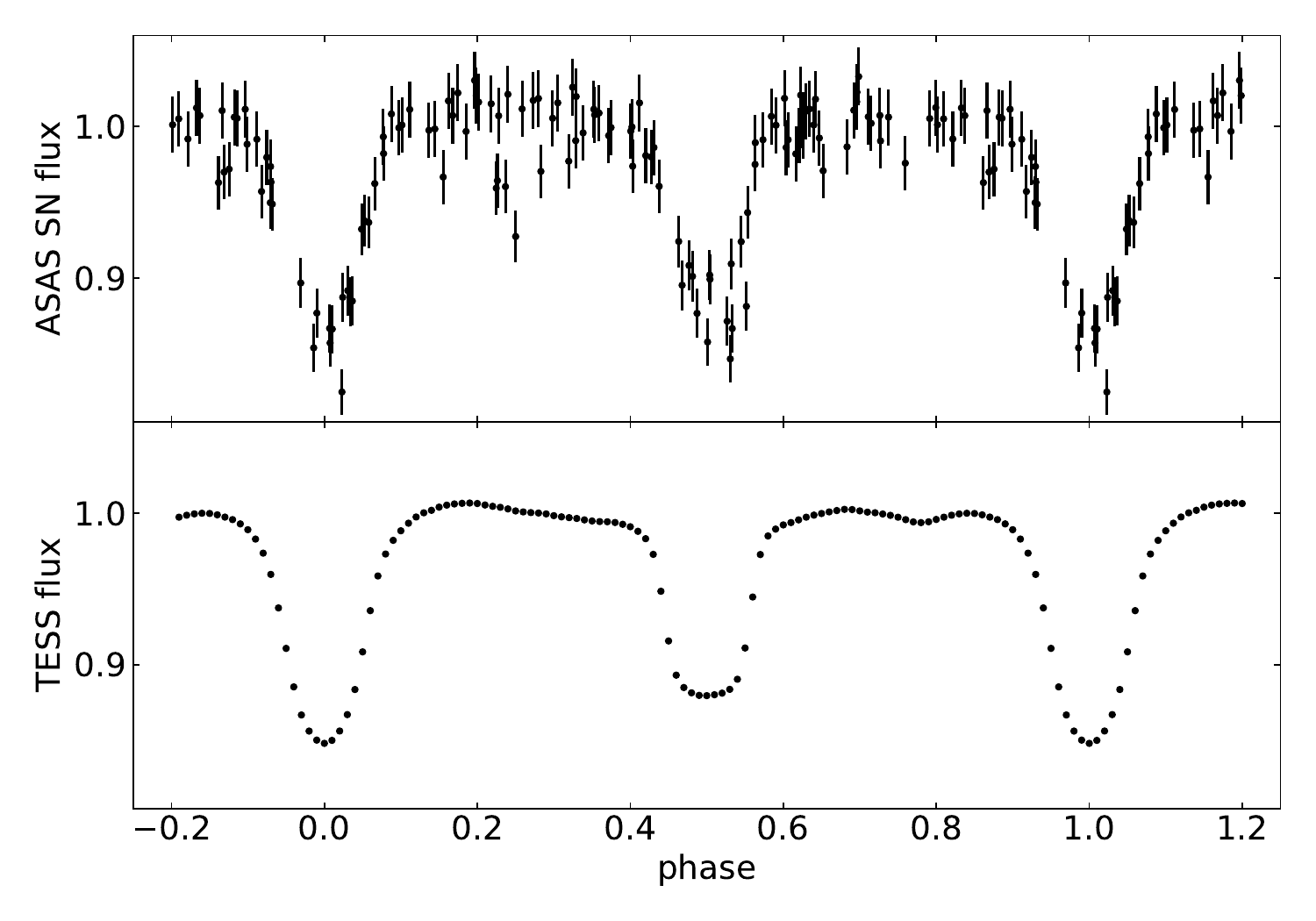}
\caption{Top panel: Normalized ASAS-SN light curve of LS\,III\,+44\,21, folded with the orbital period $P$ and the initial epoch $T_0$ from Table~\ref{tab:rv_sol2}. Bottom panel: Normalized mean TESS light curve. This curve incorporates 52\,875 individual measurements. The uncertainty of a data point on the mean TESS light curve is smaller than the symbol radius.}
\label{fig:lc_all}
\end{figure}

Figure~\ref{fig:rv_fit} shows the observational and model radial velocity curves of the LS\,III\,+44\,21 components, folded with the orbital period $P$ and the initial epoch $T_0$ from Table~\ref{tab:rv_sol2}, and shifted to the reference frame with a zero $\gamma$-velocity. Figure~\ref{fig:lc_all} displays the ASAS-SN light curve and the mean TESS light curve, both folded with the same ephemeris from Table~\ref{tab:rv_sol2}. The light curves are normalized to the mean flux level within the phase interval of $0.15-0.4$. The center of the secondary minimum occurs exactly at phase $0.5$, which is consistent with the circular orbit conclusion derived from the radial velocity curve solution.

\section{Physical Parameters of the System Components}\label{sec:phys_pars}

The parameters of the system and its components listed in Table~\ref{tab:rv_sol2} do not exhibit any extraordinary characteristics. Rather, they confirm the assumption that LS\,III\,+44\,21 is close analogue to the V444~Cyg system. For instance, the minimum orbital radius is $35.5\,R_\odot$ while the orbital size of V444~Cyg is $\sim 36\,R_\odot$. The morphology of the mean TESS light curve is highly intriguing. The wings of the secondary minimum are very steep, which unambiguously indicates that geometric, rather than atmospheric, eclipses are observed in the system, and that the eclipsed core radius of the WR star is remarkably small. This also implies that the disk of the WR star passes more or less through the center of the O star's disk. Under such conditions, one would expect to observe a total eclipse with a flat bottom. In reality, however, the bottom is rounded. We attempted to model the light curve within the standard Roche geometry framework. No combination of parameters was able to reproduce the shape of the lower portion of the secondary minimum. Furthermore, the light curve exhibits non-monotonic behavior near the quadratures (particularly around phase 0.78). Such unusual light-curve features may indicate the influence of the component stellar winds. A detailed analysis of the system's light curve is beyond the scope of this paper and will be the subject of a forthcoming study (optical and infra-red light curves, with a model of wind-wind collision). Here, we provide purely preliminary conservative estimates of the minimum orbital inclination angle. Let us assume that the eclipse of the WR star by the O star is total, but such that at the moment of conjunction, the outer circumference of the WR disk nearly touches the circumference of the O star's disk. Then,

$$
  a\cos i + R_{\rm WR} = R_{\rm O}\,,
$$

where $a$ is the orbital radius, and $R_{\rm WR}$ and $R_{\rm O}$ are the radii of the WR and O stars, respectively. Adopting typical stellar radii for these spectral types, namely $R_{\rm WR}\simeq3\,R_\odot$ \citep{hamann19} and $R_{\rm O}\simeq 10\,R_\odot$ \citep{martins05}, we obtain $a\cos i \simeq 7\,R_\odot$. Combining this with the $a\sin i$ value from Table~\ref{tab:rv_sol2} yields $i \simeq \arctan(35.5/7)=78\fd 85$ (notably, the orbital inclination in the V444~Cyg system is $78^\circ$, \citealp{ant01}). The resulting orbital radius is $a\simeq 36.2\,R_\odot$.

Adopting the derived orbital inclination angle as a first approximation yields an estimated WR star mass of $M_{\rm WR}\simeq 7.08\,M_\odot$. This value lies near the lower limit of the typical mass range for classical WR stars, which spans $5\div 50\,M_\odot$ \citep{postnov25}. At present, this represents almost the only indirect indication of the possible reason behind the lack of X-ray emission from the system. It is possible that the stellar wind of such a low-mass WR star is relatively weak, with a mass-loss rate comparable to that of the O star ($\dot{M}_{\rm O}\sim 1.0\times10^{-6}\,M_\odot\,\rm yr^{-1}$). The collision of such winds would result in relatively weak shocks and a low X-ray luminosity for the system. Furthermore, the winds of the components may undergo mutual deceleration due to the radiation field of the companion star, leading to an additional reduction in their kinetic energy. It is this kinetic energy that transforms into thermal energy of the matter within the cooling zone behind the shock fronts. The radiation of this energy is the primary source of the thermal X-ray emission.

A strong argument in favor of the above scenario for the stellar wind interaction in LS\,III\,+44\,21 and the system geometry is the spectral variability we detected in the N\,V lines. The disappearance of the N\,V lines in the spectra during the eclipse of the WR star indicates that the formation region of these lines in the WR wind is completely occulted by the disk of the O star. This is only possible for a large orbital inclination angle and a large radius of the O star (providing yet another indirect indication of an evolved luminosity class).

\begin{figure*}
\includegraphics[width=\textwidth]{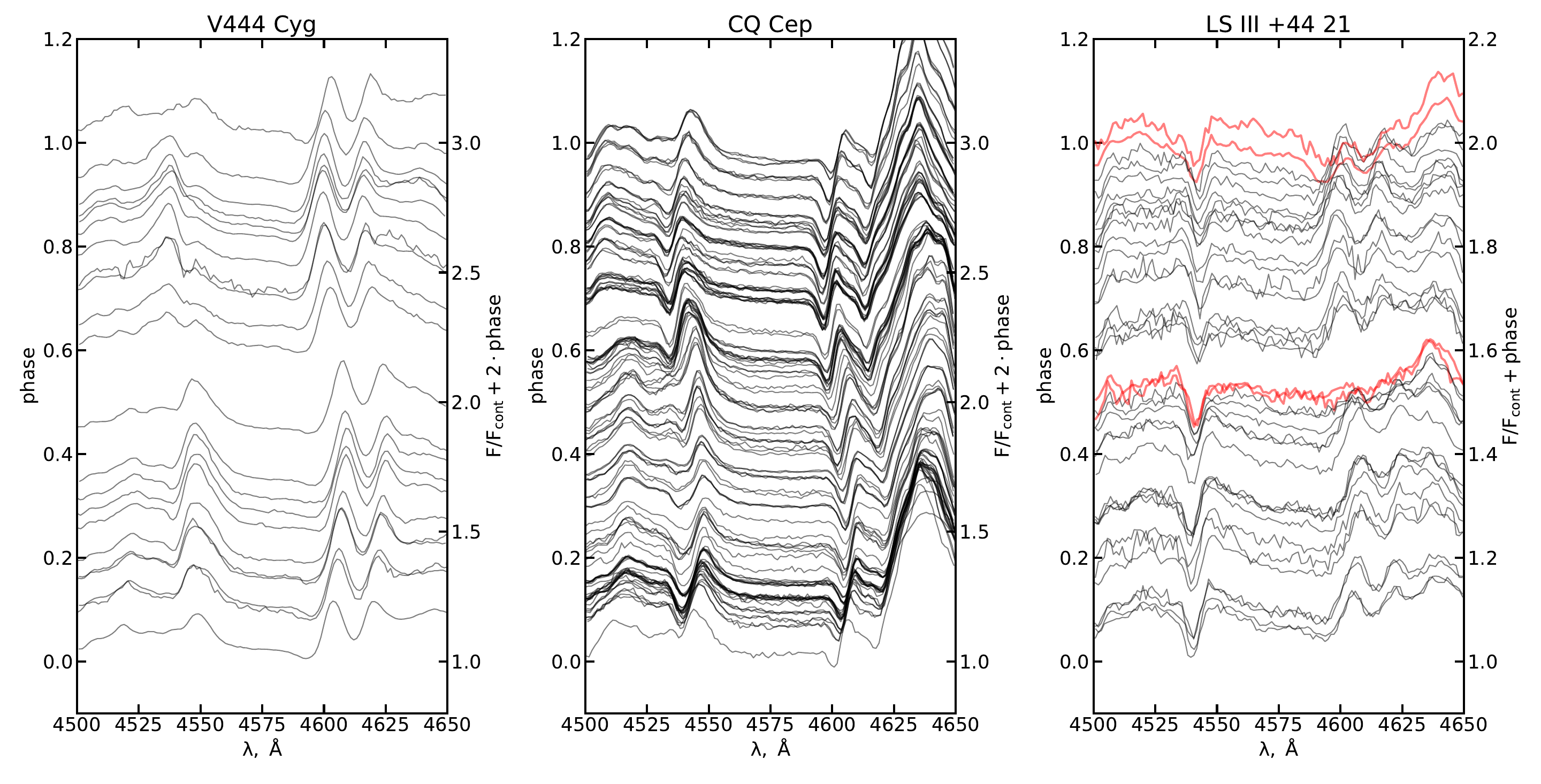}
\caption{Spectra of three close WR\,+\,OB binaries in the $4500-4650$\,\AA\ wavelength range: V444~Cyg (left), CQ~Cep (center), and LS\,III\,+44\,21 (right). The continuum level of the spectra is scaled to the orbital phase, with phase 0.5 corresponding to the eclipse of the WR star. The intensity of the V444~Cyg and CQ~Cep spectra relative to the continuum is reduced by a factor of two on the plot, while the LS\,III\,+44\,21 spectra are shown in their original scale. In the right panel, the spectra of LS\,III\,+44\,21 near phases 0(1) and 0.5, where the distortion of the \mbox{N\,V~4604-4620}\,\AA\ line profiles occurs, are highlighted in red.}
\label{fig:sp-NV4604}
\end{figure*}

The second distinctive feature of the LS\,III\,+44\,21 spectrum is the behavior of the N\,V line profiles in the opposite phase, i.e., during the transit of the WR star in front of the O star. In this case, we observe a noticeable enhancement of the absorption components in the P~Cyg profiles of the \mbox{N\,V~4604-4620\,\AA\ }lines, accompanied by a simultaneous weakening of the emission components. A similar effect has been noted previously in detailed spectroscopic studies of close WR\,+\,O binary systems; however, in the case of LS\,III\,+44\,21, it is exceptionally pronounced. Figure~\ref{fig:sp-NV4604} illustrates a portion of the LS\,III\,+44\,21 spectrum compared with the spectra of two other well-studied WR\,+\,O systems, V444~Cyg and CQ~Cep. Their spectra, like the spectra of LS\,III\,+44\,21 presented in this work, were obtained with the 2.5-meter telescope of the CMO SAI MSU using the TDS spectrograph \citep{CQ_sp23,V444_sp23}. For V444~Cyg (Figure~\ref{fig:sp-NV4604}, left panel), the observed intensity and profile variability of the \mbox{N\,V~4604-4620\,\AA\ }lines is limited to a slight strengthening of the P~Cyg absorption feature. In the case of CQ~Cep, no strong spectral variability is observed in the \mbox{N\,V~4604-4620\,\AA\ }lines either (Figure~\ref{fig:sp-NV4604}, center panel). For LS\,III\,+44\,21, however, we see a dramatic change in the appearance of this line during the eclipse phases (Figure~\ref{fig:sp-NV4604}, right panel, where the spectra in the eclipse phases are highlighted in red). Such behavior of the N\,V line profiles appears to be an indication of the prominence of the radiative inhibition and the relative weakness of the WR wind in this system. We plan to investigate this phenomenon in LS\,III\,+44\,21 in greater detail in the future, based on our forthcoming spectral observations of the eclipses in this system with a higher temporal cadence.

\section{Conclusions}\label{sec:concl}

We have carried out extensive spectroscopic observations of the recently discovered close WR\,+\,O binary system LS\,III\,+44\,21. The accumulated spectroscopic material allowed us to detect an interesting variability in the N\,IV and N\,V ion line profiles, as well as to refine the spectral classification of the system components to \mbox{WN4\,+\,O7III?(f)}.

Using these observations, we obtained the radial velocity curves of the system components for the first time. Their solution reveals that the system has a circular orbit. Incorporating photometric observations from ASAS-SN and TESS allowed us to significantly refine the values of the initial epoch $T_0$ and the orbital period $P$ of the system. The radial velocity curve solution with the updated $T_0$ and $P$ enabled the first determination of the parameters of the orbit and the system components.

The shape of the highly precise mean TESS light curve is extremely unusual, exhibiting non-monotonic behavior near the quadratures and an unusual morphology of the secondary minimum, which shows a nearly total eclipse yet with rounded edges. Such a shape cannot be modeled within the standard Roche geometry, which may provide a direct indication of the influence of the component stellar winds. Adopting the assumption of a grazing configuration of the stellar disks at the moment of the secondary eclipse as a first approximation, we estimated the orbital inclination angle and provided an initial assessment of the absolute parameters of the system and its components. The mass of the WR star is close to the lower limit of classical WR stellar masses. This indirectly suggests that the lack of significant X-ray emission from the system may be due to the low mass-loss rate of the WR star's wind.

The case of LS\,III\,+44\,21 clearly demonstrates that the initial paradigm that WR\,+\,O binaries must always act as sources of strong and hard X-ray emission is not entirely correct. The nature of stellar wind collisions in such systems is governed not only by the mass-loss rates and velocities of the winds but also, for instance, by mutual radiative inhibition in close binary systems. The WR\,+\,O binary system LS\,III\,+44\,21 serves as an excellent example of a non-standard object, the study of which can improve our understanding of wind-collision processes in such systems.

\begin{acknowledgments}

This paper includes data collected by the TESS mission, which are publicly available from the Mikulski Archive for Space Telescopes (MAST). Funding for the TESS mission is provided by NASA’s Science Mission directorate. We acknowledge the use of the data from ``All-Sky Automated Survey for Supernovae'' (ASAS-SN), which is publicly available from the ASAS-SN Photometry Database. This research has made use of data and/or software provided by the High Energy Astrophysics Science Archive Research Center (HEASARC), which is a service of the Astrophysics Science Division at NASA/GSFC.

\end{acknowledgments}

\section*{Funding}

This work was supported by the State Assignment of Lomonosov Moscow State University and partially funded by the Lomonosov Moscow State University Development Program.

\section*{Conflict of Interests}

The authors declare no conflict of interest.

\bibliographystyle{aspb1}
\bibliography{Antokhin}

\end{document}